\documentclass[prl,twocolumn,superscriptaddress]{revtex4}
\usepackage{amssymb,amsmath,amsthm,graphicx}
\usepackage{latexsym}
\usepackage{bm}
\usepackage[]{hyperref} 

\def\be{\begin{equation}}
\def\ee{\end{equation}}
\def\beq{\begin{eqnarray}}
\def\eeq{\end{eqnarray}}

\pdfoutput=1

\begin{document}

\def\lsim{\mathrel{\rlap{\lower4pt\hbox{\hskip1pt$\sim$}}
    \raise1pt\hbox{$<$}}}
\def\gsim{\mathrel{\rlap{\lower4pt\hbox{\hskip1pt$\sim$}}
    \raise1pt\hbox{$>$}}}
\def\be{\begin{equation}}
\def\ee{\end{equation}}
\def\bea{\begin{eqnarray}}
\def\eea{\end{eqnarray}}
\newcommand{\der}[2]{\frac{\partial{#1}}{\partial{#2}}}
\newcommand{\dder}[2]{\partial{}^2 #1 \over {\partial{#2}}^2}
\newcommand{\dderf}[3]{\partial{}^2 #1 \over {\partial{#2} \partial{#3}}}
\newcommand{\eq}[1]{Eq.~(\ref{eq:#1})}
\newcommand{\dd}{\mathrm{d}}

\title{Gravitational Turbulent Instability of Anti-de Sitter Space}

\author{ \'Oscar J. C. Dias}
\email{O.Dias@damtp.cam.ac.uk}
\affiliation{DAMTP, Centre for Mathematical Sciences, University of Cambridge, Wilberforce Road, Cambridge CB3 0WA, UK\\
\& Institut de Physique Th\'eorique, CEA Saclay, CNRS URA 2306, F-91191 Gif-sur-Yvette, France}
\author{Gary T. Horowitz}
\email{gary@physics.ucsb.edu}
\affiliation{Department of Physics, UCSB, Santa Barbara, CA 93106, USA.}
\author{Jorge E. Santos}
\email{jss55@physics.ucsb.edu}
\affiliation{Department of Physics, UCSB, Santa Barbara, CA 93106, USA.}

\begin{abstract}
Bizon and Rostworowski have recently suggested that anti-de Sitter spacetime might be nonlinearly unstable to transfering energy to smaller and smaller scales and eventually forming a small black hole. We consider pure gravity with a negative cosmological constant and find strong support for this idea. While one can start with a single linearized mode and add higher order corrections to construct a  nonlinear geon, this is not possible starting with a linear combination of two or more modes. One is forced to add higher frequency modes with growing amplitude. The implications of this turbulent instability for the dual field theory are discussed.
\end{abstract}

\maketitle


{\bf Introduction:} At the linearized level, anti-de Sitter (AdS) spacetime looks just as stable as Minkowski spacetime or de Sitter spacetime. There are a complete set of modes which oscillate and do not grow in time. However, at the nonlinear level, it has been suggested that AdS might behave very differently. Whereas Minkowski space and de Sitter space have been shown to be stable under small but finite perturbations \cite{Christodoulou:1993uv,Friedrich:1986}, this has never been shown for AdS. In fact the following intuitive argument suggests that it will not be the case. AdS boundary conditions act like a confining box.  Any finite excitation which is added to this box might be expected to eventually explore all configurations consistent with the conserved quantities. This includes small black holes. In other words, one might conjecture that after a sufficiently long time, any finite excitation of AdS eventually finds itself inside its Schwarzschild radius and collapses to a black hole \cite{aware}.

In a very interesting  paper  \cite{Bizon:2011gg}, Bizon and Rostworowski have recently found evidence in favor of this conjecture. They considered the spherically symmetric collapse of a massless scalar field in AdS.  No matter how small they made the initial amplitude for the scalar field, their numerical evolution always produced a black hole. They also did a perturbative analysis of the problem and found a possible loophole in the above argument. If they started with a single linearized mode of the scalar field in AdS, they could add nonlinear corrections systematically in a way which suggested that no singularity will form. However, if they  started with a general superposition of linearized modes and tried to add higher order corrections, they had to add higher frequency modes. They also showed that under evolution, the energy in the initial modes decrease while the energy in the higher frequency modes grow. They argued that this is analogous to a turbulent instability in which energy is transferred from larger to smaller scales. This suggests that a generic finite perturbation of AdS will indeed eventually collapse to a black hole.

One drawback of their analysis was that it was restricted to spherical symmetry, so gravitational degrees of freedom were not excited. In addition,  one might wonder if an angular momentum barrier will prevent a similar collapse at late times. We study the purely gravitational problem of the nonlinear stability of AdS and find results very similar to \cite{Bizon:2011gg}. Starting with a single linearized mode, we compute higher order perturbative corrections and find no obstruction.   However as soon as one has a superposition of linearized gravitational modes,  higher order corrections force one to add higher frequency modes with growing amplitudes.  Although we do not follow the evolution to see whether a black hole eventually forms, the following argument suggests that it will. One of the singularity theorems shows that closed universes are generically singular \cite{Hawking:1969sw}. This theorem does not apply to AdS directly (e.g. because spacelike surfaces are not compact) but morally speaking, the negative cosmological constant acts like a confining box for fields inside. So one expects that generic solutions will be singular.

There is no contradiction between this nonlinear instability of AdS and the positive energy theorem. The latter says that if you start with AdS, it will not decay or evolve to anything else since it is the unique solution with zero energy. The question of whether small amounts of energy can collapse to small black holes is very different and usually ruled out by arguing that waves disperse at late time. This does not happen in AdS.

The nonlinear generalizations of individual perturbative modes are {\it geons}, i.e., lumps of gravitational energy which are held together by their own self gravity.  They are  nonsingular, asymptotically globally AdS,  and can be viewed as  gravitational analogs of boson stars. Since we can start with any linearized mode, there are an infinite number of geons. However, these solutions are all special in that they each have one (helical) Killing field. They are exactly periodic in time (answering a question raised in \cite{Anderson:2006ax}). The geons are stable to linearized perturbations, but will be unstable at higher order to the turbulent instability.  We will argue that one can put small rotating black holes inside the geons and obtain black holes with only a single Killing field. (These are purely gravitational analogs of the black holes constructed in \cite{Dias:2011at}.) Thus, Kerr AdS is not the only stationary, asymptotically AdS black hole. This had previously been argued for based on the superradiant instability of Kerr AdS.  Our small black holes sitting inside large geons provide another class of examples.

Physically, one can view the turbulent instability as a result of one geon focussing the energy of the other. It is perhaps analogous to the collision of plane gravitational waves which results in singularities in finite time due to similar focussing. Of course one difference is that  plane waves produce singularities after one collision. In our case, it takes many collisions of the geons before one expects singularities to form.


{\bf Perturbation theory:}
We shall focus on the four-dimensional case \cite{higherD}, described by the following action:
\begin{equation}
S=\int d^4x\;\sqrt{-g}\left(R+\frac{6}{L^2}\right),
\label{eq:action}
\end{equation}
where $L$ is the AdS length scale. We perturb the equations of motion derived from (\ref{eq:action}) by expanding the metric about the AdS background, \emph{i.e.} $g = \bar{g}+\sum_i h^{(i)}\epsilon^i$, where $\epsilon$ is  a perturbation parameter whose physical meaning will be discussed later, and $\bar{g}$ is the metric of AdS written in global coordinates
\begin{equation}
\bar{g} = -\left(1+\frac{r^2}{L^2}\right)dt^2+\frac{dr^2}{1+\frac{r^2}{L^2}}+r^2(d\theta^2+\sin^2\theta d\phi^2).
\end{equation}
At each order in perturbation theory, the Einstein equations yield
\begin{equation}
\Delta_L h_{ab}^{(i)} = T^{(i)}_{ab},
\label{eq:perturb}
\end{equation}
where $T^{(i)}$ is a function of $\{h^{(j\leq i-1)}\}$ and their derivatives and $\Delta_L$ is a second order operator constructed from $\bar{g}$,
\begin{equation}
2\Delta_L h_{ab}^{(i)} \equiv -\bar \nabla^2 h_{ab}^{(i)}-2 \bar{R}_{a\phantom{c}b\phantom{d}}^{\phantom{a}c\phantom{b}d}h_{cd}^{(i)}-\bar{\nabla}_{a}\bar{\nabla}_b h^{(i)}+2\bar{\nabla}_{(a} \bar{\nabla}^c h^{(i)}_{b)c}.
\end{equation}
Here, $h^{(i)}\equiv \bar{g}^{ab}h_{ab}^{(i)}$, and $\bar{R}_{abcd}$ is the AdS Riemann tensor. At second order $T^{(2)}$ reduces to the familiar Landau-Lifshitz pseudotensor \cite{Landau:1951}. As a consequence of the Bianchi identities, $\bar \nabla^a T^{(i)}_{ab} =0$ for each $i$.

We will construct regular finite energy and angular momentum solutions of (\ref{eq:perturb}) for $i \geq 1$. In \cite{Kodama:2003jz}, the $SO(3)$ symmetry of AdS was used to show that any regular two-tensor, say $A$, can be expressed as an infinite sum of two simple building blocks
\begin{equation}
A = \sum_{\ell_s,m_s} A^{(s)}_{\ell_s,m_s}+\sum_{\ell_v,m_v} A^{(v)}_{\ell_v,m_v}+\cos \phi\leftrightarrow\sin \phi,
\label{eq:linearsol1}
\end{equation}
where $A^{(s)}$ ($A^{(v)}$) represent scalar (vector) type modes, which are symmetric two-tensors built from scalar (vector) harmonics on the two-sphere. Here $(\ell,m)$ are the usual labels for spherical harmonics, and vector harmonics on $S^2$ are of the form ${}^*\nabla Y_{\ell,m}$.  The last term in (\ref{eq:linearsol1}) accounts for the fact that we are using a real representation for the spherical harmonics, so both the $\sin \phi$ and $\cos \phi$ terms need to be included. Each $A^{(s)}_{\ell_s,m_s}$  ($A^{(v)}_{\ell_v,m_v}$) is parametrized by seven  (three) arbitrary functions of $t$ and $r$.

It is possible to show, by applying the expansion (\ref{eq:linearsol1}) to both $h^{(i)}$ and $T^{(i)}$ in (\ref{eq:perturb}), that any solution of (\ref{eq:perturb}) is described by two decoupled PDEs of the form
\begin{equation}
\Box_s \Phi^{(i)}_{\ell,m}(t,r)+V^{(i)}_{\ell}(r) \Phi^{(i)}_{\ell,m}(t,r)=\tilde{T}^{(i)}_{\ell,m}(t,r),
\label{eq:master}
\end{equation}
where $\Phi^{(i)}_{\ell,m}(t,r)$ is a Kodama-Ishibashi like (gauge invariant) variable from which  $h^{(i)}_{\ell,m}$ can be recovered (in a particular gauge)  through a linear differential map \cite{Kodama:2003jz}. One of the equations governs vector-type modes, while the other is related to scalar-type modes. Regular solutions of (\ref{eq:master}) are in one to one correspondence with smooth solutions of (\ref{eq:perturb}). Here, $\tilde{T}^{(i)}_{\ell,m}(t,r)$ is a scalar source term, that can be expressed as a function of the components of $T^{(i)}_{\ell,m}$ and its derivatives, $V^{(i)}_\ell(r)$ is a potential that can be found in \cite{Kodama:2003jz} and $\Box_s$ is the d'Alambertian associated with the auxiliary orbit space $ds^2 = -(1+r^2/L^2) dt^2+L^2 dr^2/(L^2+r^2)$.

At each order in perturbation theory, we will impose regularity of $h^{(i)}_{\ell,m}$, seen as a tensor on a fixed AdS background. This in turn induces very strong regularity conditions on any solution of (\ref{eq:master}). In order for $h^{(i)}_{\ell,m}$ to have a regular centre, $\Phi^{(i)}_{\ell,m}\sim \mathcal{O}(r^{\ell})$, as $r\to0$. At asymptotic infinity, we require the metric to be asymptotically globally AdS. The asymptotic form of $\tilde{T}^{(i)}_{\ell,m}$ is such that any solution of (\ref{eq:master}) behaves as
\begin{equation}
\Phi^{(i)}_{\ell,m}\sim R_{\ell,m}(t)+ \frac{S_{\ell,m}(t)}{r}+\mathcal{O}(r^{-2}),
\label{eq:asymp}
\end{equation}
where $R_{\ell,m}$ and $S_{\ell,m}$ are arbitrary functions of $t$. Note that the condition of asymptotically globally AdS is imposed on $h^{(i)}_{\ell,m}$  \emph{not} on $\Phi^{(i)}_{\ell,m}$ directly. This means that we must reconstruct $h^{(i)}_{\ell,m}$ before imposing the boundary conditions on $\Phi^{(i)}_{\ell,m}$. It turns out that if we want our geometry to approach the Einstein static universe at the boundary, we must keep the leading term and choose $S_{\ell,m}(t)=0$.

A couple of comments about Eq.~(\ref{eq:master}) are in order: i) we need to compute all the previous $h^{(i)}_{\ell,m}$ in order to calculate the source term in (\ref{eq:master}) at the next order; ii) if the time dependence of the source term is an overall multiplicative factor of the form $\cos(\tilde{\omega}_{\ell,m} t)$, (\ref{eq:master}) can be solved by assuming a separable form for $\Phi^{(i)}_{\ell,m}(t,r)=\cos(\tilde{\omega}_{\ell,m} t) R(r)$. The only  exception to (ii) arises if and only if $\tilde{\omega}_{\ell,m}$ coincides with any of the AdS gravitational normal frequencies described below. In this case, the modes are said to be \emph{resonant}, and the general solution of (\ref{eq:master}) now grows with time:
\begin{multline}
\Phi^{(i)}_{\ell,m}(t,r)=\cos(\tilde{\omega}_{\ell,m}t)R_{\tilde{\omega}_{\ell,m}\ell,m}(r)\\+t\,\sin(\tilde{\omega}_{\ell,m}t)L_{\tilde{\omega}_{\ell,m},\ell,m}(r).
\label{eq:resonant}
\end{multline}

At first order, $\tilde{T}^{(1)}_{\ell,m}\equiv0$, and one can Fourier transform Eq.~(\ref{eq:master}) in time, with Fourier parameter $\omega_\ell$, reducing the problem to the study of a single ODE of the Sturm-Liouville type. For simplicity, we focus on scalar-type modes. Requiring a regular centre and the solution to be asymptotically globally AdS imposes $L\omega_\ell = 1+\ell +2 p$, where $p\in\{0,1,2\ldots\}$ is the radial overtone. The positivity of $\omega^2_\ell$ indicates that AdS is linearly stable. For $p=0$, (\ref{eq:master}) admits a simple solution:
\begin{equation}
\Phi^{(1)}_{\ell,m}(t,r) = \frac{r^{\ell+1}}{(r^2+L^2)^{\frac{\ell+1}{2}}} a_{\ell,m}\,\cos(\omega_\ell\,t),
\label{eq:solution}
\end{equation}
where $a_{\ell,m}$ is a normalization constant. Gravitational waves correspond to $\ell \ge 2$.

Following \cite{Bizon:2011gg}, we have chosen two types of initial data: one is just $\ell =2, m = 2$ and the other is a linear combination of $\ell =2, m = 2$ and $\ell =4, m = 4$.
For the single mode initial data, we choose the normalization and the phase so that 
\be\label{linearized}
(2 L^2h^{(1)}_{tt}/r^2)|_{r=0} = -9 \sin ^2\theta\cos \left(\frac{3 t}{L }-2 \phi \right).
\ee
 We then calculate $T^{(2)}$, and rewrite it as in (\ref{eq:linearsol1}), with a total of six independent terms, corresponding to six values for the pair $(\ell,m)$. This means we have to solve for a total of six PDEs of the form (\ref{eq:master}). The solutions can be found and rendered regular both at asymptotic infinity and at the centre of AdS. At third order we find that $T^{(3)}$ can also be expressed as a sum of six terms, five of which behave as the second order terms, but one, with $\ell_s = m_s =2$ and $L\,\tilde{\omega}_{2,2}=3$, is resonant. This could lead to the secular behavior exhibited in (\ref{eq:resonant}). However it turns out that we can set $L_{3,2,2}(r)=0$ by promoting the frequency of our initial data to be a function of $\epsilon^2$: $L\,\omega_2 = 3-14703\,\epsilon^2/17920 $. Thus, to third order, the solution is regular everywhere, with no growing modes in time. It is invariant under a Killing vector which is a slightly shifted version of the symmetry of the linearized mode (\ref{linearized}):  $K\equiv\partial_t+\frac{\omega_2}{2}\partial_\phi$. We have computed $T^{(4)}$, and calculated the energy as a function of the angular momentum to fifth order in $\epsilon$:
\begin{equation}\label{geonEJ}
E_g = \frac{3J_g}{2L}\left(1-\frac{4901\,J_g}{7560\pi L^2}\right),\quad \omega_2 = \frac{3}{L}\left(1-\frac{4901\,J_g}{3780\pi L^2}\right),
\end{equation}
where we defined $\epsilon$ by $J_g = \frac{27}{128}\pi L^2\epsilon^2$. It can also be checked that this solution obeys the first law of thermodynamics $dE_{g}=\frac{\omega}{m} \,dJ_{g}$ to $\mathcal{O}(J^{3}_{g})$. From the structure of the equations, we expect the situation at higher orders to be similar. It is easy to show that there are no resonant terms at any even order. We expect  one resonant mode at each odd order, but by adjusting the frequency one can eliminate the growing mode. The result is a (nonlinear) geon.  

The two mode initial data behaves quite differently.  At the linear level we start with a composite mode $\Phi^{(1)}_{2,2}(t,r)+\Phi^{(1)}_{4,4}(t,r)$. At second order we find that $T^{(2)}$ is now expanded as a sum of 17 terms, none of which are resonant. For each of these terms one can calculate the right hand side of (\ref{eq:master}) and a regular solution can always be found. The situation changes at third order. Here we find that $T^{(3)}$ can be expressed as a sum of 36 harmonics, 32 of which behave just as the second order perturbations. The remaining four correspond to resonant modes. Two of the resonant modes, with $L\,\tilde{\omega}_{2,2}=3,\,m_s = \ell_s=2$ and $L\,\tilde{\omega}_{4,4}=5,\,m_s = \ell_s=4$, can be made regular at infinity and the origin by a suitable change in the frequencies of the initial data, just like we did for the single mode case. For the resonant mode with $L\,\tilde{\omega}_{0,0} =1,\,m_s=\ell_s=0$ the solution is already regular with $L_{1,0,0}(r)=0$ in (\ref{eq:resonant}). The interesting behavior occurs for the resonant mode with the highest possible frequency, $L\,\tilde{\omega}_{6,6} = 7,\,m_s = \ell_s = 6$. In this case the solution can only be made regular if $L_{7,6,6}(r)$ in (\ref{eq:resonant}) is non-zero, leading to a power law growth in time of the initial perturbation, and thus to an instability. It is significant that although there are four resonant modes at third order, only the highest frequency one leads to a growing mode. This is the expected behavior for a turbulent flow, for which the amplitude of higher frequency modes should become larger as time passes by. We expect that when this term is large enough, it will create other resonant modes with higher frequency and that this process will end with the formation of a rotating black hole.

{\bf Black holes with only one Killing field:}
We now argue that one can place a small Kerr AdS black hole in the core of the geon without disturbing it.  Since the geon has only one Killing field,  the resulting black hole will have only a single (helical) Killing field. The argument is based on earlier  studies where explicit examples of scalar hairy black holes were  constructed. When the black holes are small, their leading order thermodynamics is accurately reproduced by a non-interacting mixture of two components.  In one study, the components were charged black holes and charged solitons \cite{Basu:2010uz}. In another, they were rotating black holes and rotating boson stars \cite{Dias:2011at}. Similarly, we assume that a small black hole does not interact with a large geon.  Absence of interaction means that the charges ${E,\,J}$ of the final black hole are simply the sum of the charges of its individual constituents:  $E=E_{K}+E_{g}$, $J=J_{K}+J_{g}$. The Kerr component controls the entropy and the temperature of the final black hole: the geon has no entropy and has undefined temperature. Since the geon has only one Killing field and we place a Kerr black hole with a Killing horizon at its center, the geon's Killing field must coincide with the horizon generator of the black hole. That is, the angular velocity of the horizon must be  $\Omega_H=\frac{\omega}{m}$. This thermodynamic equilibrium condition also follows from maximizing the entropy for a given total energy and angular momentum.
Combining this condition with the thermodynamics of the two components, it follows that at leading order, the geon carries all the angular momentum of the system.

Another way to construct black holes with only a single Killing field is through the superradiant instability of Kerr AdS black holes with $L\,\Omega_H > 1$ \cite{Kunduri:2006qa}. This instability causes certain modes to grow outside the horizon, extracting energy and angular momentum from the black hole. If you perturb a black hole by a single unstable mode, then it will grow and eventually settle down to a black hole with ``gravitational hair" outside the horizon. For small black holes, the onset of this instability (for a mode with azimuthal quantum number $m$) can be determined by simply setting $\Omega_H = \frac{\omega}{m}$, where $\omega$ is the frequency of the linearized geon. (It suffices to consider the linearized geon since the ``hair" is very weak near the onset of the instability.) This condition is motivated both by the Killing field argument above and the fact that it saturates the condition for a superradiant instability: $\omega \leq m \Omega_H$. One finds 
\begin{equation}\label{onset}
 E|_{onset}\simeq  \frac{ r_+}{2}+\frac{r_+^3}{2L^2} \left(1+\frac{\omega^2L ^2}{m^2}\right), \quad J|_{onset}\simeq\frac{1}{2}r_+^3\frac{\omega}{m}
 \end{equation} 
 This simple argument reproduces the known results for  black holes with scalar hair.  In \cite{Dias:2011at} it was shown that in a phase diagram of $E$ vs $J$,  black holes with scalar hair exist in the region below the onset of superradiance and above
the rotating boson star curve. Similarly, we expect that vacuum black holes with a single Killing field exist between  (\ref{onset}) and the geon curve
(\ref{geonEJ}).


{\bf Discussion:}
There is a connection between the turbulent instability  and the  superradiant instability. Superradiance causes modes to grow. If you perturb a black hole by a single unstable mode, then it will eventually settle down to a hairy black hole.  However,  if you perturb the black hole with two unstable modes, both modes will start to grow. Their interaction will trigger the turbulent instability which will transfer energy to higher and higher frequency. 

Our discussion so far has been entirely classical. Quantum mechanically, we expect that the transfer of energy from large to small scales will still occur, but now will be cut off at a frequency equal to the initial energy. So if one starts with an initial state consisting of a large number of low energy quanta with total energy less than the Planck energy, one will never form a black hole.

We now comment on the dual field theory interpretation of our results using gauge/gravity duality. Since we have only considered gravity in the bulk, any field theory with a gravity dual must exhibit the same turbulent instability, and transfer energy from large to small scales. The fact that this is so universal seems surprising, although the final outcome of a small black hole can be viewed as thermalization in a microcanonical ensemble. It is even more surprising  considering the fact that our dual system is  $2+1$ dimensional, and in this case classical turbulence causes energy to flow from small to large scales \cite{GT}. Our results indicate that (at least at large $N$) strongly coupled $2+1$ dimensional quantum theories in finite volume behave very differently. We have found similar results in AdS$_5$, which might appear to contradict the idea that the large $N$ limit of super Yang-Mills is integrable. However the discussion of integrability does not usually include states with energy of order $N^2$ needed for finite backreaction.

Our nonlinear geons also have a dual interpretation. Since the field theory is on a sphere, at zero temperature it is in a confined phase in the sense that the free energy is order one and not a power of $N$. A linearized graviton can be viewed as a spin two excitation (e.g. a glue-ball). The nonlinear geon is dual to a bose condensate of these excitations. It is interesting that  (at large $N$ and strong coupling) these high energy states do not thermalize.

We have argued that asymptotically AdS solutions are generically singular. It would be of great interest to prove a singularity theorem establishing this rigorously.

\vskip .5cm
\centerline{\bf Acknowledgements}
\vskip .2cm
We wish to thank E. Martinec, D. Marolf and J. Polchinski for discussions. This work was supported in part by the  National Science Foundation under Grant No.~PHY08-55415.
OJCD acknowledges financial support from an European Marie Curie contract.



\begin{thebibliography}{99}
\bibitem{higherD}
  We have performed analogous calculations in five dimensions, and the results are qualitatively similar.
\end{thebibliography}
\end{document}